\documentclass[12pt]{article}
\usepackage{graphicx} 
\usepackage{amsmath,caption,threeparttable} 

\begin{document}
	
\title{{\bf Two-mirror aplanatic telescopes \\ with a flat field}} 
\author{V.~Yu.~Terebizh\thanks{E-mail: valery@terebizh.ru}
\thanks{Experimental Astronomy DOI: 10.1007/s10686-022-09833-0}\\
\small{Crimean Astrophysical Observatory,} \\ 
\small{Nauchny, Crimea 298409, Ukraine}}
	
 \date{\small{{\it February 10, 2022}}} 
	
\maketitle 

\begin{abstract}
A complete description is given of two-mirror telescopes with a flat medial 
focal surface, on which the images of stars are {\it circles of least 
confusion}. Particular attention is paid to aplanats, since their field of 
view is noticeably larger than that of classical systems. Two sets of 
appropriate solutions correspond to Schwarzschild and Gregorian telescopes. 
As a result, it becomes possible to use flat light detectors with wide-field 
two-mirror telescopes. New designs are of particular interest when as few 
reflective surfaces as possible are required, which is typical for space 
exploration and non-optical observations. 
\end{abstract}

\section{Introduction} 

By definition, {\it aplanatic} optical systems are those in which spherical 
aberration and coma are corrected. For a pure mirror aplanat, the field of 
view is limited primarily by astigmatism, curvature of the focal surface, 
and, to a lesser extent, by distortion. The combined action of astigmatism 
and field curvature results in two focal surfaces of paraboloidal shape, 
{\it tangential} and {\it sagittal}, with the paraxial radii of curvature 
$R_t$ and $R_s$, on which the images of stars are elongated in mutually 
perpendicular directions (see, e.g., Hecht~[1], p.~264). Generally speaking, 
further correction of astigmatism does not ensure a flat focal surface: the 
tangential and sagittal surfaces merge with each other, forming a single 
{\it Petzval surface} with the finite paraxial radius of curvature $R_P$, 
and only the addition of the {\it Petzval condition} provides a flat shape 
of the focal surface (Born and Wolf~[2], Section 5.5.3; Korsch~[3], 
Section~6.5). 

The complete definition of the two-mirror system contains too few initial 
parameters, namely two, to correct astigmatism and field curvature, so the 
main instruments of the 20th century were two-mirror aplanats, which include 
the Schwarzschild, Ritchey-Chr\'etien systems and Gregory-Maksutov mirror 
aplanats (Schroeder~[4], Terebizh~[5]). In particular, the Hubble Space 
Telescope (HST) is a Ritchey-Chr\'etien aplanat with radii of curvature 
$R_t = -574.730$~mm, $R_s = -693.941$~mm, and $R_P = -774.237$~mm. 
All image surfaces here are concave with respect to the direction of the 
light rays, and the Petzval surface has the smallest curvature. 

\begin{figure}[ht] 
	\centering  
	\includegraphics[width=65mm]{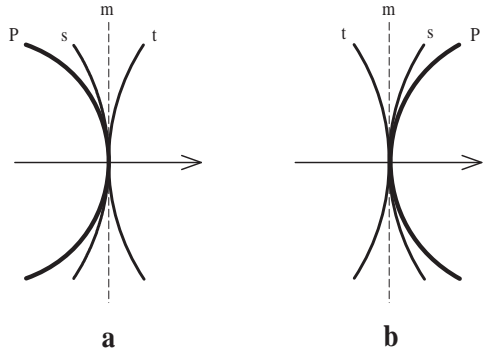} 
	\caption{{\small 
	Two types of image surfaces allowing a flat medial surface. 
	P, s, m and t denotes the Petzval, sagittal, medial, and tangential 
	focal surfaces, respectively.}} 
\end{figure} 

Since flat light detectors still prevail, two-mirror telescopes with flat 
image surfaces, be they classical systems or aplanats, are of undoubted 
interest. Some special cases of aplanatic two-mirror telescopes with a flat 
field are known (for example, Schwarzschild exact aplanat), but this 
problem has not been systematically considered. Such a study can be carried 
out with the help of the procedure called {\it artificial flattening} 
(Hecht~[1], p.~269). It consists in the arrangement of the tangential and 
sagittal surfaces symmetrically relative to the focal plane, and then the 
{\it medial} surface lying between them will be flat (Figure~1). Of course, 
images of stars on the medial surface ({\it circles of least confusion}) 
are still spoiled by astigmatism, but they are the smallest and round. Pay 
attention: it is desirable to make flat not the notional Petzval surface, 
but precisely the medial surface. 

The purpose of this paper is to give a simple and complete description of 
two-mirror telescopes that combine the wide field of view of aplanats with 
the possibility of using flat radiation detectors.

\section{Arbitrary two-mirror telescope} 
\label{Any} 

Let us recall two results obtained by Joseph Petzval in the middle of the 
19th century (Born and Wolf~[2]). The first is the relation 
$$ 
\frac{3}{R_s} - \frac{1}{R_t} = \frac{2}{R_P} 
\eqno(1) 
$$
that is valid for an arbitrary optical system.\footnote{Equation~(1) is 
in the wrong place in the original version of the Experimental Astronomy 
article.} Besides, the tangential and sagittal surfaces are always on the 
same side of the Petzval surface. The second result is a special case of 
the general representation of the Petzval radius for a two-mirror system: 
$$
\frac{1}{R_P} = -\frac{2}{R_1} + \frac{2}{R_2}\,, 
\eqno(2) 
$$
where $R_1$ and $R_2$ are paraxial radii of curvature of the primary 
and secondary mirrors, respectively. 

In general, the medial surface is specified by the condition 
$$
\frac{1}{R_m} \equiv \frac{1}{2}\left(\frac{1}{R_t}+\frac{1}{R_s}\right), 
\eqno(3)
$$
which requires its curvature to be the arithmetic mean of the curvature 
of the tangential and sagittal surfaces. 

For what follows, it is necessary to find explicit representations of all 
four radii of curvature in terms of the parameters of the optical system. 

\begin{figure}[h] 
	\centering  
	\includegraphics[width=120mm]{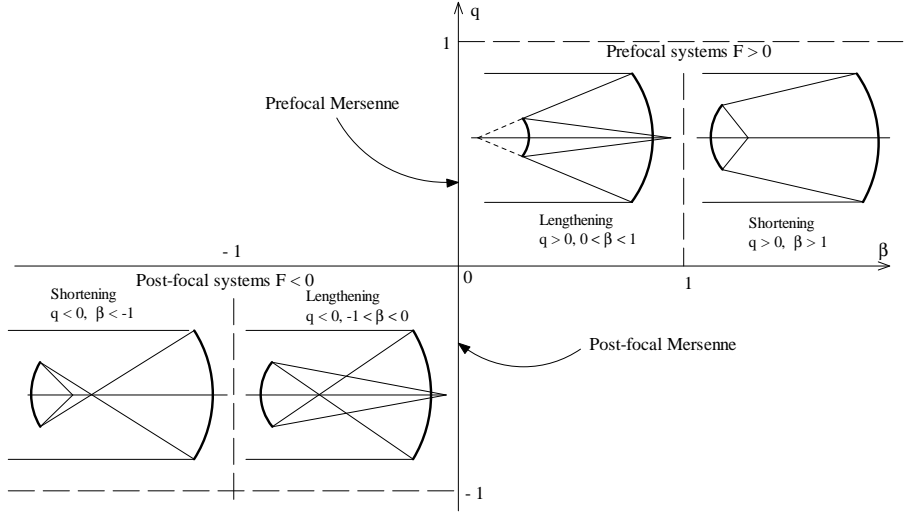} 
	\caption{\small{
	Schematic representation of two-mirror telescopes on the Maksutov 
	diagram. First and third quadrants correspond to systems in which 
	secondary mirror is located, respectively, before the primary focus 
	and behind it; two other quadrants correspond to systems with virtual 
	images. Further subdivision of systems is determined by whether the 
	focal length of the primary mirror is lengthened or shortened by the 
	secondary mirror. Mersenne systems correspond to the 
	$\beta = 0$\, axis.}} 
\end{figure}

Various sets of initial parameters are used to describe two-mirror 
telescopes; for both the optical systems and their aberrations we use the 
simple description by Maksutov~[6,7]. Extension of this approach to arbitrary 
two-mirror telescopes is illustrated in Figure~2 (Terebizh~[5]). The scheme 
comes from four parameters: the telescope aperture diameter $D$, the 
equivalent  focal length $F$, and two dimensionless ratios 
$$
q = s_2^{'}/F, \quad \mbox{and} \quad  \beta = f_1/F,  
\eqno(4)
$$
where $s_2^{'} > 0$ is the back focal length of the telescope, that is the 
distance from the secondary mirror to the axial focal point, and $f_1$ is the 
focal length of the primary mirror. As usually, we assume $f_1 = -R_1/2 > 0$ 
for a concave primary mirror. The value of $F$ can be either positive or 
negative.\footnote{Recall that the effective focal length $F$ of an optical 
system is measured from the back principal plane to the focal plane. 
Therefore, the negativity of~$F$, as is the case for the Gregorian system, 
means that the back focal plane is located farther behind the focus along the 
optical axis. To facilitate the use of various sources, we give the 
relationship between the Schroeder~[4] and Maksutov~[6,7] variables: 
$m_{Schr} = 1/\beta = m$, $\beta_{Schr} = q(1+1/\beta)-1$.} Let $D_{2}^{(0)}$ 
be the secondary mirror light diameter for the axial incident light beam; 
then $|q| = D_{2}^{(0)}/D \le 1$ is the linear obstruction ratio for the 
above beam, whereas $\beta$ is the inverse magnification $m = F/f_1$ of the 
secondary  mirror. Parameters $q$ and $\beta$ have the same sign for 
telescopes with real images. As Figure~2 shows, we have positive $F$ for the 
Cassegrain and Schwarzschild systems, and negative $F$ for the Gregorian 
telescopes. 

In Maksutov variables, $R_1 = -2\beta F$, and $R_2 = qR_1/(1-\beta)$, so 
equation~(2) gives the Petzval radius 
$$
R_P = -\frac{q\beta F}{1-q-\beta}\,. 
\eqno(5) 
$$
This expression is true for all two-mirror telescopes, whereas the form of 
$R_t$ and $R_s$ depends on the specific type of system.

\section{Two-mirror aplanat}
\label{Aplanat}

When considering aplanatic two-mirror telescopes, we proceeded from the 
general theory of third-order aberrations, as expressed by Maksutov~[6] and 
Mikhelson~[8]. Rather cumbersome calculations, which we omit here, lead to 
the following expressions for the paraxial radii of curvature of the image 
surfaces, namely, the tangential surface: 
$$
R_t = -\frac{2q\beta F}{4\beta-(1-q)(3\beta^2-2)}\,, 
\eqno(6)
$$
and the sagittal surface: 
$$
R_s = -\frac{2q\beta F}{(1-q)(2-\beta^2)}\,. 
\eqno(7)
$$  
It is easy to check that the last three formulas agree with the general 
Petzval relation~(1). Further, it follows from equations~(3), (6) and (7) 
that 
$$
R_m = -\frac{q\beta F}{\beta+(1-q)(1-\beta^2)}  
\eqno(8)
$$
for aplanatic telescopes. Referring again to the HST example with its 
effective focal length $F = 57599.869$~mm and parameters $q = 0.111219$, 
and $\beta = 0.095834$, we get $R_m = -628.735$~mm from equation~(8). 

\begin{figure}[h] 
	\centering  
	\includegraphics[width=100mm]{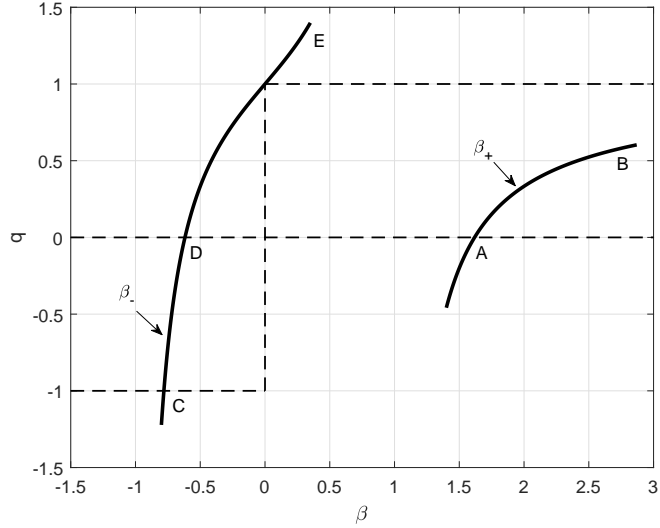} 
	\caption{\small{
		Bold lines in the Maskutov diagram correspond to the solutions 
		$\beta_{+}$ and $\beta_{-}$ with a flat medial surface. Only the 
		$AB$ branch and the $CD$ section are in the zones of physically 
		acceptable optical systems, bounded by the dotted lines.}} 
\end{figure} 

As stated in the Introduction, by requiring $R_t = -R_s$ we arrive at one 
of the symmetrical arrangements of the tangential and sagittal surfaces 
relative to the medial surface, shown in Figure 1. The same result 
follows from~(8) by equating the denominator to zero. The corresponding 
aplanatic systems with a flat medial surface are defined in the plane 
$(q,\beta)$ by equation 
$$
q = 1+\frac{\beta}{1-\beta^2}\,, 
\eqno(9)
$$
or its solutions 
$$
\beta_{\pm}(q) = \frac{1\pm \sqrt{1+4(1-q)^2}}{2(1-q)} 
\eqno(10) 
$$ 
in the form of a quadratic equation with respect to $\beta$ (Figure~3).
Comparison of Figures~2 and~3 shows that the first of these solutions 
corresponds to the shortening Schwarzschild aplanat in the interval 
$\beta \ge (1+\sqrt{5})/2 \simeq 1.618$ (branch AB in Figure~3), whereas 
the second one~-- to the lengthening Gregory-Maksutov aplanat in the 
interval $(1-\sqrt{17})/4 \le \beta \le (1-\sqrt{5})/2$, that is 
approximately $-0.781 \le \beta \le -0.618$ (section CD in Figure~3). 

Note that both branches AB and CD imply $R_2 > 0$, i.e. concave shape of 
the secondary mirror. Since $R_1 < 0$ for the primary mirror, it follows 
from~(2) that $R_P > 0$ for the two-mirror aplanats with a flat medial 
surface.  Thus, the curvatures of both branches are represented by the 
option `b' in Figure~1.

\begin{figure} [ht]
	\begin{center} 
		\begin{tabular}{cc}
			\includegraphics[width=65mm]{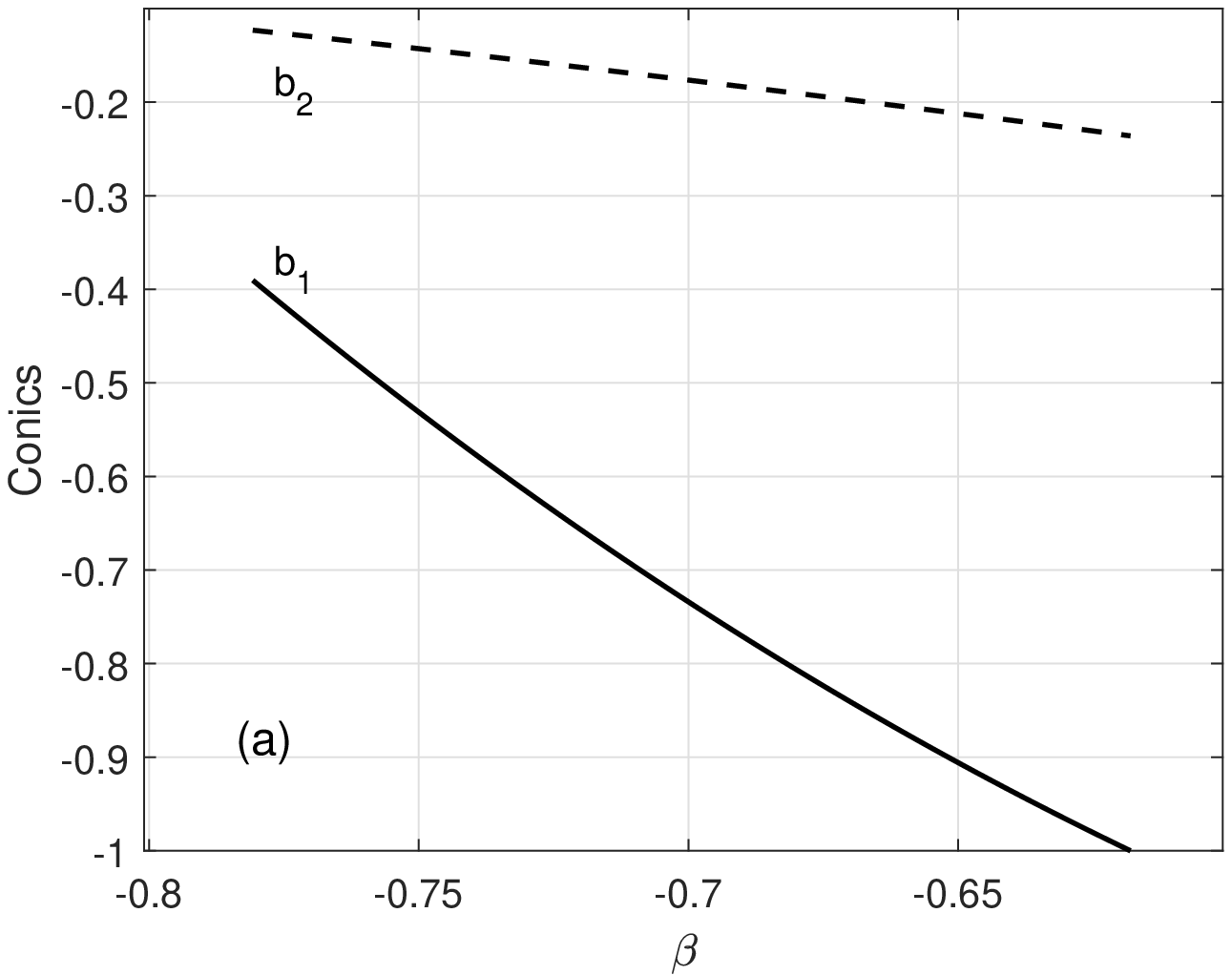} & 
			\includegraphics[width=65mm]{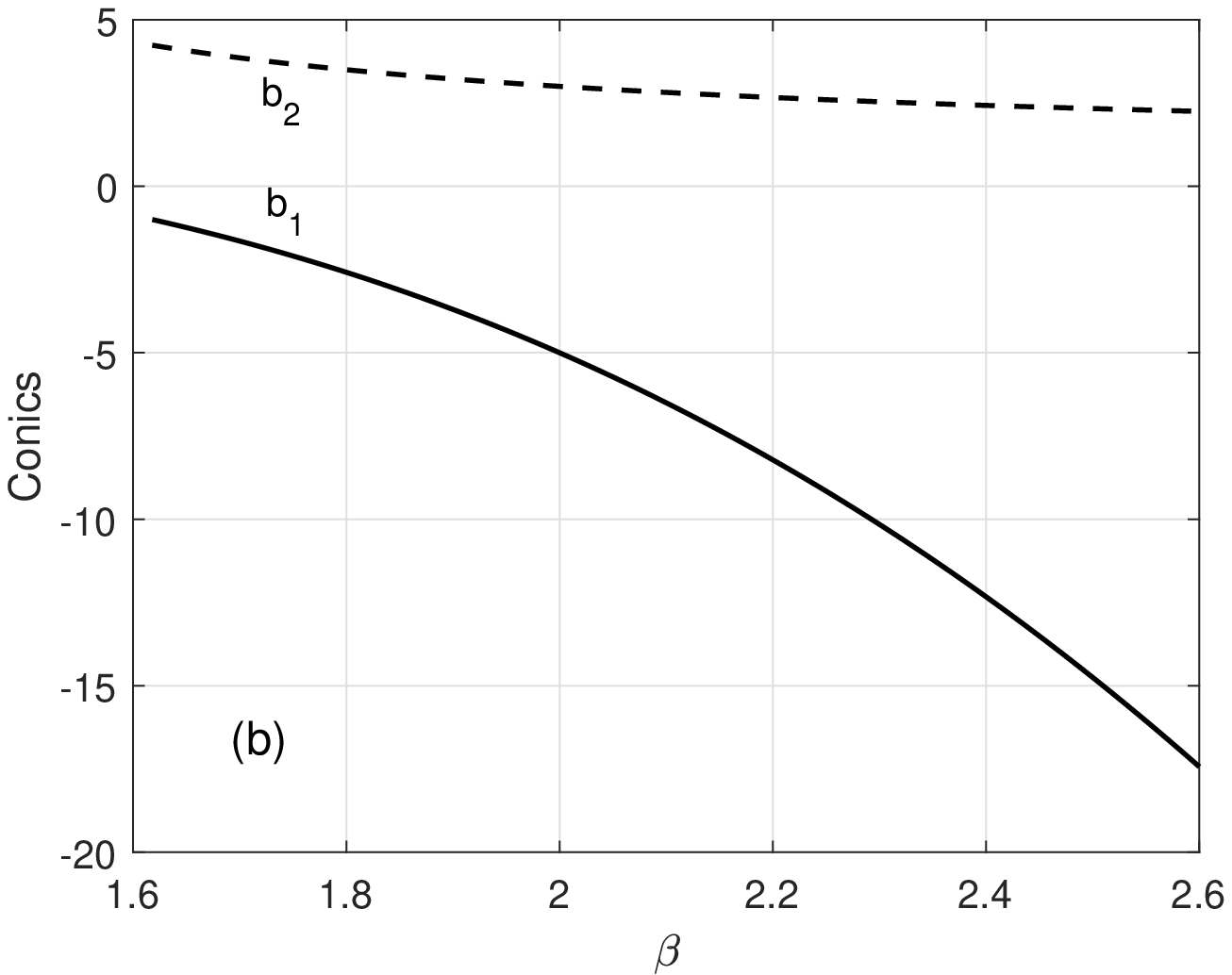} \\ 
		\end{tabular} 
		\caption[f04]
		{\label{fig:f04} 
		\small{Conic constants for the two-mirror telescopes with a flat 
		medial surface.\\
		Figure~(a) corresponds to systems with the negative effective 
		focal length~$F$,\\ Figure~(b)~--  to systems with positive~$F$.}} 
	\end{center} 
\end{figure} 

The above results provide a complete description of the sought systems, but 
it is useful to add that relations~(9) and (10) reduce the two-parameter 
representation of telescopes in the plane $(q,\beta)$ to a one-parameter 
representation of flat-field aplanats, either by means of $q$ or by means 
of $\beta$. Equations (2.6)~-- (2.13) of Terebizh~[5] concerning the 
first-order characteristics of two-mirror systems are simplified accordingly. 
If we choose $\beta$ as the initial parameter, then the obscuration 
coefficient $q$ is given by equation~(9), the back focal length $s_2' = qF$, 
the radius of curvature of the primary mirror $R_1 = -2\beta F$, the same 
parameter for the secondary mirror 
$$
R_2 = -\frac{2\beta F}{1-\beta}\left(1+\frac{\beta}{1-\beta^2} \right),
\eqno(11)
$$
and the spacing between mirrors 
$$
T = \frac{\beta^2 F}{1-\beta^2}. 
\eqno(12)
$$
As for the conic constants of mirrors, equations~(2.16) of the mentioned 
book take the form for flat-field aplanats: 
$$
b_1 = -1 + 2\beta(1+\beta-\beta^2), \qquad 
b_2 = -\frac{1+\beta}{1-\beta}\,. 
\eqno(13)
$$
Last equations are presented graphically in Figure~4. We see that at 
negative values of $F$, both mirrors have the shape of prolate ellipsoids 
of revolution. This was to be expected, since the mirrors of all aplanats 
in the region ($q < 0;\, -1 < \beta < 0$) have this shape. At $F > 0$, 
that is, in the area of the Schwarzschild aplanats, the primary mirror is 
a hyperboloid, while the secondary is an oblate ellipsoid.

\begin{table}[ht] 
	\begin{center} 
	\captionsetup{font=small} 
	\caption{Two designs with different signs of~$F$.} 
	\label{tab:1} 
	\begin{threeparttable} 
		\begin{tabular}{cll} 
			\hline\noalign{\smallskip} 
			Parameter   & Design A      & Design B    \\
			\noalign{\smallskip}\hline\noalign{\smallskip}
			$D$   		& $1000.0$      & $1000.0$        \\ 
			$F$   		& $3000.0$      & $-3000.0$      \\
			$\beta$ 	& $2.00$        & $-0.70$      \\
			---         & ---------     & ---------     \\
			$q$         & $1/3$  		& $-0.372549$   \\
			$R_1$       & $-12000.0$ 	& $-4200.0$  \\
			$R_2$       & $4000.0$  	& $920.415225$  \\
			$T$         & $-4000.0$ 	& $-2882.352941$  \\
			$s_2^{'}$   & $1000.0$ 		& $1117.647059$  \\
			$b_1$       & $-5.0$  		& $-0.734000$   \\
			$b_2$       & $3.0$  		& $-0.176471$   \\
			$R_t$       & $-3000.0$     & $-754.966887$   \\
			$R_s$       & $3000.0$  	& $754.966887$   \\
			$R_P$       & $1500.0$  	& $377.483444$   \\
			\noalign{\smallskip} 
			\hline 
		\end{tabular} 
		\begin{tablenotes} \footnotesize 
			\item[Note] Linear dimensions are given in millimeters. 
			The input parameters are separated by a line. $T$ means 
			the distance between the mirrors, $s_2^{'}$~-- the 
			back focal length. 
		\end{tablenotes} 
	\end{threeparttable} 
	\end{center}
\end{table}

\section{Examples of aplanatic designs} 
\label{Examples}

Let us consider two examples corresponding to branches $\beta_{+}(q)$ and 
$\beta_{-}(q)$ of solutions with a flat medial surface. 

For the Design~A, we took the aperture diameter $D = 1.0$~m, the effective 
focal length $F = 3.0$~m, and $\beta = 2.0$, which, according to~(9), 
corresponds to $q = 1/3$. These data are sufficient to calculate all the 
other parameters of the telescope; they are shown in the second column 
of Table~1. Entering numerical values into the ZEMAX program shows 
that the images of point sources are round throughout the entire flat 
field of view of the order of 20~arc minutes in diameter. 

Similar data for the Design~B are shown in the third column of Table~1. 
In this case, we adopted the same aperture, but the negative effective 
focal length $F = -3.0$~m, and $\beta = -0.70$. Again, the images of the 
stars are round in the flat field of view with a diameter greater than 
$1^\circ$. Another thing is that the images at the edges of the field 
become blurred, but their symmetrical shape indicates the flatness of the 
focal surface.

\section{Classical telescopes}

Systems with a flat medial surface exist also among {\it classical} 
two-mirror telescopes, that is, telescopes in which only spherical 
aberration is corrected. We pay less attention to these systems, since 
coma noticeably limits their field of view. 

Within the framework of the theory of third-order aberrations, the radii 
of curvature of the tangential and sagittal image surfaces are now: 
$$
R_t = -\frac{q\beta F}{2\beta+(1-q)(1-3\beta^2)}\,, 
\eqno(14)
$$
$$
R_s = -\frac{q\beta F}{(1-q)(1-\beta^2)}\,. 
\eqno(15)
$$
Taking into account~(3), we obtain the radius of curvature of the 
medial surface: 
$$
R_m = -\frac{q\beta F}{\beta+(1-q)(1-2\beta^2)}\,. 
\eqno(16)
$$
The Petzval radius $R_P$ is still given by~(5). 

Like before, symmetrizing the arrangement of the tangential and sagittal 
surfaces in the form $R_t = -R_s$ or zeroing the denominator in~(16) 
leads to the condition of flatness of the medial surface: 
$$
q = 1+\frac{\beta}{1-2\beta^2}\,, 
\eqno(17)
$$
and its solutions 
$$
\beta_{\pm}(q) = \frac{1\pm \sqrt{1+8(1-q)^2}}{4(1-q)} 
\eqno(18) 
$$ 
as a quadratic equation with respect to $\beta$. This time there are also 
two branches of physical solutions: the first in the interval $\beta > 1$ 
within the first quadrant of the $(q,\beta)$ plane, and the second branch 
in the interval $-1/\sqrt{2} < \beta < -1/2$ within the third quadrant 
of this plane (see Figure~2). Otherwise, the results are similar to those 
for aplanatic systems.

\section{Concluding remarks} 
\label{Remarks}

The above analytical calculations and numerical examples discuss the 
options for the existence of two-mirror telescopes with round images of 
stars on a flat detector of light within the field of view typical of 
aplanats. Unfortunately, the set of these systems is not abundant. The 
reason lies in the small number of free parameters that describe two-mirror 
telescopes. In this respect, greater opportunities provide three-mirror 
telescopes, for which it is possible to correct in addition both astigmatism 
and field curvature, and thereby achieve a true {\it anastigmatic} system. 
Corresponding algorithms for calculating three-mirror anastigmats were 
proposed by Korsch~[3] and Terebizh~[5,9]. 

Despite the relative scarcity of flat-field two-mirror models, it is 
likely that they may be of interest for astrophysical and astrometric 
studies when as few reflective surfaces as possible are required, which 
is typical for space exploration and off-optical observations.

\subsection*{Acknowledgments}

I am grateful to the anonymous Referee for helpful comments.

\end{document}